# Why Kastner analysis does not apply to a modified Afshar experiment

Eduardo Flores and Ernst Knoesel

Department of Physics & Astronomy, Rowan University, Glassboro, NJ 08028

In an analysis of the Afshar experiment R.E. Kastner points out that the selection system used in this experiment randomly separates the photons that go to the detectors, and therefore no which-way information is obtained. In this paper we present a modified but equivalent version of the Afshar experiment that does not contain a selection device. The double-slit is replaced by two separate coherent laser beams that overlap under a small angle. At the intersection of the beams an interference pattern can be inferred in a non-perturbative manner, which confirms the existence of a superposition state. In the far field the beams separate without the use of a lens system. Momentum conservation warranties that which-way information is preserved. We also propose an alternative sequence of Stern-Gerlach devices that represents a close analogue to the Afshar experimental set up.

## 1. INTRODUCTION

The Afshar experiment is a version of a double slit experiment first suggested and carried out by Afshar.[1] Recently Afshar et al.[2] reported on a more detailed experiment that claims to have simultaneously determined complementary wave and particle aspects of light beyond the limitations set by Bohr's Principle of Complementarity. Researchers aware of the important implications of Afshar's claims have analyzed the experiment and have either come forth or against it.[3,4] A particularly important paper against Afshar's claims was written by R.E. Kastner.[5] In her analysis of Afshar's experiment Kastner draws an analogy with a typical spin measurement of a spin-½ particle, and based on this comparison she concluded that Afshar's claim of full "which-way" information is unjustified.

In this paper we first review the Afshar experiment and its analysis by Kastner. We propose an alternative set up made of ideal Stern-Gerlach (SG) devices that interact with spin-½ particles in a similar way the Afshar experimental set up interacts with photons. We then present a modified but equivalent version of Afshar's experiment that utilizes momentum conservation to determine the trajectory of the particle.

## 2. THE AFSHAR EXPERIMENT

In the Afshar experiment[2] coherent light is incident onto a pair of pinholes (see Fig.1). The two emerging beams from the pinholes spatially overlap in the far-field and interfere to produce a pattern of alternating light and dark fringes. At an appropriate distance from the pinholes thin wires are placed at the minima of the interference pattern.

Beyond the wires there is a lens that forms the image of the pinholes onto two photon detectors located at the image of each pinhole.

When an interference pattern is not present, as in the case when only one pinhole is open, the wire grid obstructs the beam and produces scattering, thus reducing the total flux at the detectors. However, when the interference pattern is present the disturbance to the incoming beam due to the wires is minimal. From comparative measurements of the total flux with and without the wire grid, the presence of an interference pattern is inferred in a non-perturbative manner. Thus, the parameter *V* that measures the visibility of the interference pattern is near its maximum value of 1.

When the wire grid is not present quantum optics predicts that a photon that hits detector 1' (2') originates from pinhole 1 (2) with a very high probability. The parameter *K* that measures the "which-way" information is 1 in this case. When a wire grid is placed at the dark fringes, where the wave-function is zero, the photon flux at the detectors hardly changes. Afshar[1,2] argues that this is an indication that the wires have not altered the "which-way" information, thus, *K* is nearly 1.

A modern version of Bohr's principle of complementarity is known as the Greenberger-YaSin inequality $V^2 + K^2 \leq 1$.[6] Since the values of the visibility and the "which-way" information are both close to 1 the Greenberger-YaSin inequality is grossly violated in the Afshar experiment.

### 3. KASTNER'S ANALYSIS OF THE AFSHAR EXPERIMENT

Kastner considers a Hilbert space spanned by two basis vectors $|1\rangle$ and $|2\rangle$. She then defines the superposition state $|S\rangle$, with

$$|S\rangle = \frac{1}{\sqrt{2}}[|1\rangle + |2\rangle]. \qquad (1)$$

Applied to the Afshar experiment, the state $|1\rangle$ corresponds to the state of a photon when pinhole 1 is open and state $|2\rangle$ is the corresponding state for the case when pinhole 2 is open. The state $|S\rangle$ corresponds to both pinholes open.

The Afshar experiment consists of preparing the state $|S\rangle$ at a time $t_0$. This is done by letting coherent light go through both pinholes. At a later time $t_1$ a measurement takes place by the mere presence of the wire grid. Kastner argues that the measurement that takes place at the wire grid is a confirmation measurement. The photon approaches the wire grid in state $|S\rangle$. Since the wire grid is at a place where $|S\rangle$ has zero probability the photon does not interact with it and passes the wire grid is the same state $|S\rangle$. Had the photon reached the grid in state $|1\rangle$ or $|2\rangle$ it would have experience measurable diffraction. Finally, the lens and detectors serve to provide a sharp measurement at $t_2$ either at detector 1' or 2'. According to Kastner,[4] the lens decomposes the state $|S\rangle$ randomly into state $|1\rangle$ 50% of the time and $|2\rangle$ 50% of the time. Her analysis implies that a click at a certain detector does not allow concluding through which pinhole the photon went. According to this analysis the visibility of the pattern as measured by the wire grid is nearly 1 but the "which-way" information is zero and the Greenberger-YaSin inequality is fulfilled at its limit, $V^2 + K^2 = 1$.

Kastner reaches the above conclusions by drawing an analogy of the Afshar experiment with a typical spin measurement of a spin-½ particle with respect to orthogonal spatial directions, say **x** and **z** (see Fig. 2a). Suppose a spin-½ particle comes out of an x-oriented Stern-Gerlach (SG) device in a spin up state $|x\uparrow\rangle$, which could be written as

$$|x\uparrow\rangle = \frac{1}{\sqrt{2}}[|z\uparrow\rangle + |z\downarrow\rangle]. \tag{2}$$

Kastner argues that this is mathematically equivalent to the state of a photon that goes through two open pinholes at time $t_0$. The measurement of the state of the photon that takes place at the wire grid is a confirmation measurement and corresponds to inserting an additional x-oriented SG device at time $t_1$. Finally, at time $t_2$, a z-oriented SG device gives information about the particle spin along **z**. This last measurement has an equal probability of finding the particle with spin up or down. But more importantly, we cannot draw any conclusion about the spin of the particle along the z-axis at the time $t_0$. Kastner[5] maintains that: "The lens serves to provide for a sharp measurement of the outcome either $U$ (up along **z**) or $L$ (down along **z**) at $t_2$." Thus the last z-oriented SG device plays the role of the lens in the Afshar experiment.

## 4.  STERN-GERLACH SEQUENCE FOR THE AFSHAR EXPERIMENT

In the following we want to discuss an alternative analogue to the Afshar experiment represented by four ideal SG devices (see Fig. 2b). The initial beam is obtained by passing an electron beam through an x-oriented SG device blocked at the $|x\downarrow\rangle$ exit. The beam that goes through in state $|x\uparrow\rangle = \frac{1}{\sqrt{2}}[|z\uparrow\rangle + |z\downarrow\rangle]$ is passed through a second z-oriented SG device. This second z-oriented SG device corresponds to the pinholes in the Afshar experiment. The two beams that exit, $|z\uparrow\rangle$ and $|z\downarrow\rangle$, spread and mix. In the region where the beams mix interference takes place. A third x-oriented SG device with the $|x\downarrow\rangle$ exit blocked is placed in this region. The blocked exit is not an obstacle since the electron in this region is in the interference state of $|x\uparrow\rangle$. The purpose of the blocked device is only to show that interference is present; it plays the role of the wires in the wire-grid of the Afshar experiment. Finally the mixed beam is once again separated by a z-oriented fourth SG device; the two emerging beams end at two corresponding detectors. This fourth z-oriented SG device plays the role of the imaging lens in the Afshar experiment.

We may remove the third x-oriented SG device (the wire-grid). In this case if the detector located at the $|z\downarrow\rangle$ exit of the fourth z-oriented SG device (imaging lens) clicks it means that the electron came from the $|z\downarrow\rangle$ exit of the second z-oriented SG device (the pinholes). Thus, we have full which-way information as in the Afshar experiment with no wire-grid. When we place the third x-oriented SG device (the wire-grid) in the region where the beams mix and interfere we expect that every electron that enters will exit unchanged. From comparative measurements of the total flux with and without the

third x-oriented SG device (wire-grid), the presence of an interference pattern is inferred. Thus, the parameter *V* that measures the visibility of the interference pattern is near its maximum value of 1. Since the third x-oriented SG device (wire-grid) is not expected to affect the beam, we can also assume that the which-way information is also near 1.

We note that if the beams that come out of the second z-oriented SG device (the pinholes) are directed in such a way that they separate in the far field after crossing (see Fig. 3) we would not need the fourth z-oriented SG device (imaging lens). This last set up corresponds to the modified Afshar experiment as described below.

5.   MODIFIED AFSHAR EXPERIMENT

What makes the Afshar experiment so interesting is the possibility that one can trace back the photon from the detector, say 1', to its corresponding pinhole 1. A similar treatment is possible for photons at detector 2'. This technique is simple and requires only the application of an imaging lens. Bartell[7] and Wheeler[8] have previously discussed the use of imaging lenses for obtaining which-way information in double-slit type experiments. However, it is not crucial to use a lens system to obtain which-way information. In principle, an equivalent experiment could be done without the use of a lens by employing two coherent beams that intersect at a small angle. Therefore, the modified Afshar experiment, described in this section, allows a far more transparent analysis of the mechanism that protects the "which-way" information. A version of this experiment was suggested by Wheeler.[8] Using Wheeler's version, Afshar reported[9] measurements in agreement with previous results of the Afshar experiment[2].

As seen in Fig. 4, a laser beam impinges on a 50:50 beam splitter and produces two spatially separated coherent beams of equal intensity. The beams overlap at some distance, where they form an interference pattern of bright and dark fringes. At the center of the dark fringes we place thin wires. Note that similar set-ups are commonly utilized in interference lithography, where pulsed lasers are used to create 2D and 3D periodic structures for gratings and photonic crystals. The patterns are formed by evaporating material at the position of the bright fringes of the interference pattern. Beyond the region of overlap the two beams fully separate again. There, two detectors are positioned such that detector 1' detects only the photons originating from mirror 1, and detector 2' detects only photons originating from the beam splitter (mirror 2). Since the pathway of the photon is practically unobstructed, a study of the electric fields involved[2] together with conservation of momentum[9] allows us to uniquely identify, with high probability, the respective mirror as the place where that photon originated. Thus, momentum conservation allows us to claim that the "which-way" parameter *K* is close to 1. The fringe visibility *V* can be experimentally estimated in a similar manner as in the classic Afshar experiment, in which a value of *V > .64* was obtained.[2] Thus, the Greenberger-YaSin inequality is also violated in the modified Afshar experiment.

It is interesting to analyze this experiment as Kastner suggested. Assume that the beam intensity is low enough that only a single photon is present throughout the apparatus as in the Afshar experiment.[2] At time $t_0$ the photon is in a superposition state $|S\rangle = \frac{1}{\sqrt{2}}[|1\rangle + |2\rangle]$. At a latter time $t_1$ the photon reaches the wire grid and goes through it unperturbed confirming that it is in state $|S\rangle$ at a region where there is interference. From the wire grid the photon continues to propagate unaltered until it finally reaches at time $t_2$, for example, detector 1'. As the detector 1' clicks, conservation of momentum indicates that this photon originated from mirror 1 with high probability. Consequently, we have a situation where the wave-function of the photon is in a superposition, all the way from mirrors at time $t_0$ to the wire grid at $t_1$ to the detectors at $t_2$, yet, depending on of what detector triggers, the path of the photon can be uniquely traced back to the time $t_0$, as if the photon has moved as a point particle in a straight line (see Fig. 5). We have a paradox similar to the one proposed in the Afshar experiment.[2]

The resolution of the paradox is simple if one allows the possibility that wave and particle properties can coexist. Quantum mechanics allows for the possibility that at time $t_0$ the photon wavefunction is in the superposition state $|S\rangle$ while the actual particle is at mirror 1. In fact the mathematical framework of quantum mechanics does not contain any information on the actual location of a point particle such as the photon or the electron; it deals with the wave aspect only.

## 6.    CONCLUSION

In response to Kastner's analysis we propose an alternate Stern-Gerlach sequence, which represent an analogue to the Afshar experiment. In addition we introduce a modified Afshar experiment, in which the beams separate by themselves without the necessity of an imaging lens. As a result the photon is undisturbed and remains in the same superposition state all the way from mirrors to detectors. Yet, when either detector clicks, the path of the photon can be uniquely traced by applying the law of conservation of momentum. Thus, a simple quantum mechanical analysis of the modified version of the Afshar experiment results in a paradox. The resolution of the paradox appears to imply that complementarity does not apply in cases where the visibility of the interference pattern can be determined in a non-perturbative way. This modified Afshar experiment clearly reveals the coexistence of particle and wave properties as if they were two separate entities.

7. ACKNOWLEDGEMENTS

The authors are grateful with S. S. Afshar for sharing with them his experiment and with Prof. Tony Heinz of Columbia University for fruitful discussions and for pointing out to them the equivalence between the Afshar experiment and the crossed beam experiment.

**FIGURES**

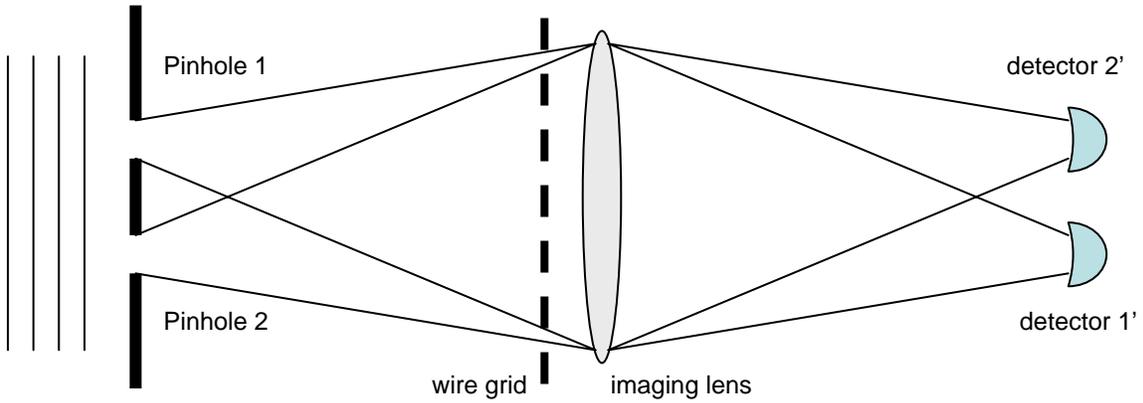

Fig 1: The Afshar Experiment

b. The Afshar Experiment for ½-spin Matter

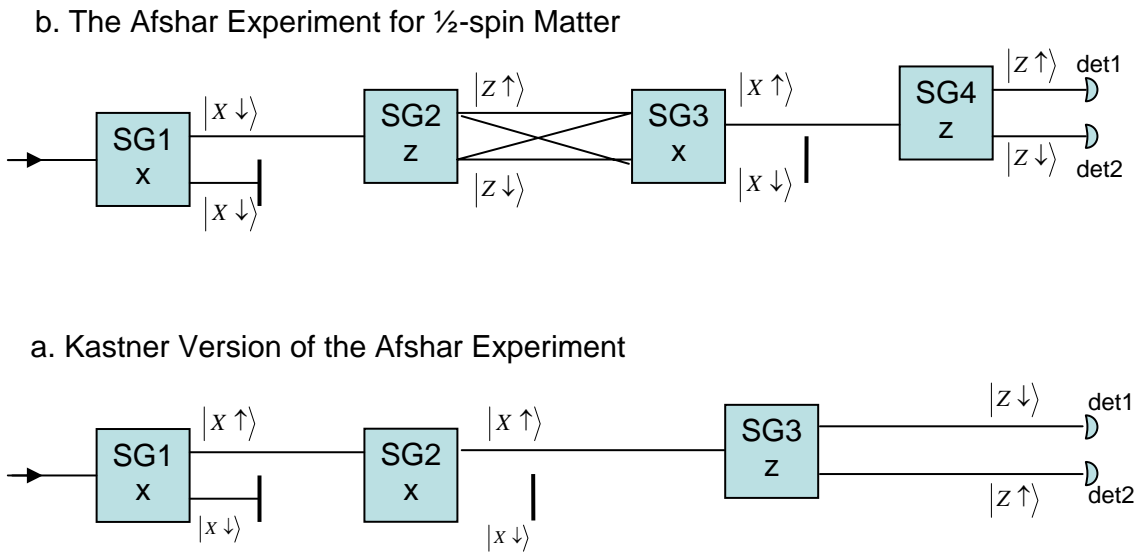

a. Kastner Version of the Afshar Experiment

Fig 2: a) Kastner Version of the Afshar experiment. The x-SG1 represents the pinholes. The x-SG2 is the equivalent of the wire-grid in the Afshar experiment. Here a confirmation measurement of the x-up superposition state occurs. The z-SG3 plays the role of the lens in the Afshar experiment and separates the beam into their respective z-state components. b) Sequence of Stern-Gerlach (SG) set-ups as an analogue for the Afshar experiment for Matter. The x-SG1 represents the step to create a coherent beam of ½-spin particles oriented in the x-up direction. The z-SG2 represents the pinholes. At the x-SG3 the beams overlap and a confirmation measurement of the x-up superposition state occurs. The z-SG4 plays the role of the lens in the Afshar experiment and separates the beam into their respective z-state components. $|x\downarrow\rangle$

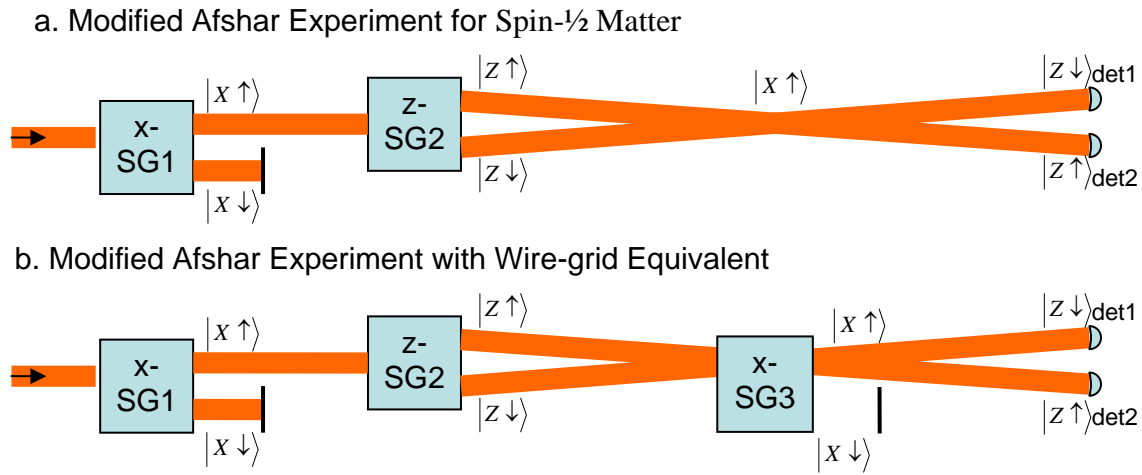

Fig 3: a) Sequence of Stern-Gerlach (SG) set-ups for Modified Afshar experiment for spin-½ matter. The x-SG1 represents the step to create a coherent beam of ½-spin particles oriented in the x-up direction. The z-SG2 represents the pinholes. The outgoing beams are directed at an angle; where the beams cross interference takes place and the state $|x\uparrow\rangle$ is formed. In the far field the beams separate again and end at corresponding

detectors. b) Modified Afshar experiment with wire-grid equivalent (x-SG3). At the x-SG3 the beams overlap and a confirmation measurement of the x-up superposition state occurs.

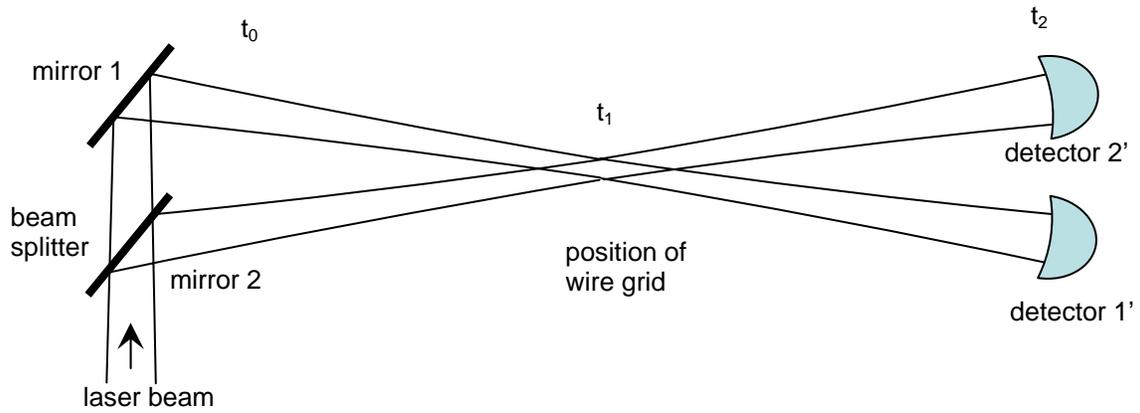

Fig 4. Modified Afshar experiment; the separation of the two beams occurs without an imaging system.

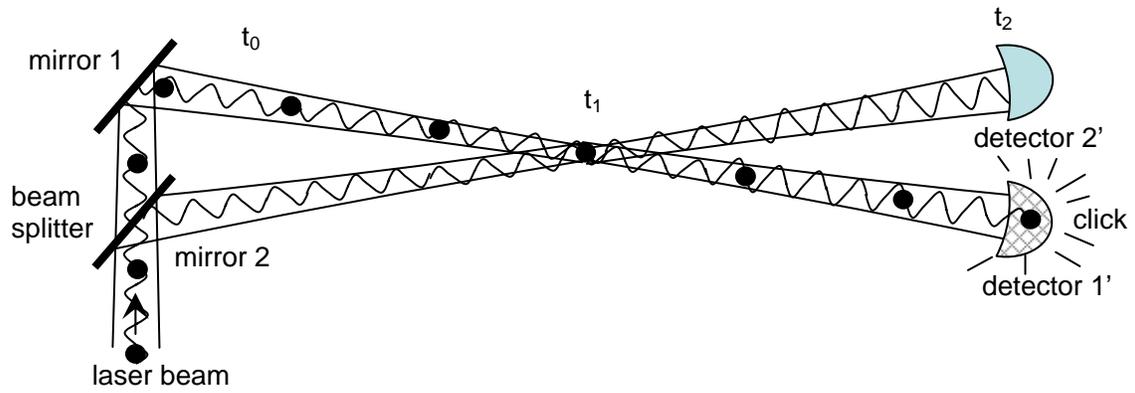

Fig 5. A photon enters a modified Afshar experiment in a superposition state, which is denoted by waves. Once the photon is detected in detector 1' its trajectory can be traced back to mirror 1 due to momentum conservation (black dots).